\documentclass[aps,preprint,groupedaddress]{revtex4}
\usepackage{epsfig,amsbsy,bm}
\usepackage{graphicx}
\usepackage{dcolumn}
\usepackage{bm}

\newcommand{\eref}[1] {(\ref{#1})}
\newcommand{\Eref}[1] {Eq.~(\ref{#1})}

\begin{document}

\baselineskip = 8mm

\title{Helium atom in the monochromatic electromagnetic field: \\
a Hylleraas basis treatment}


\title{Calculations of total photoionization cross-section for
two-electron atomic systems}

\author{I. A. Ivanov\footnote[1]{Corresponding author:
Igor.Ivanov@.anu.edu.au}\footnote[2]{On leave from the Institute of
Spectroscopy, Russian Academy of Sciences} and A. S. Kheifets}
\affiliation{Research School of Physical Sciences and Engineering,
The Australian National University,
Canberra ACT 0200, Australia}

\date{\today}

\begin{abstract}
We outline a non-perturbative procedure for calculating the total
photoionization cross-section of two-electron atomic systems. The
procedure is based on the Floquet-Fourie representation of the
solution of the time-dependent Schrodinger equation. The
Floquet-Fourie ansatz produces a set of equations which is recast into
a generalized eigenvalue problem by means of the complex rotation
procedure. With the use of the Hylleraas-type basis functions, the
total photoionization cross-sections are obtained within the accuracy
of a fraction of a percent.  The total photoionization cross-sections
for neutral helium are in good agreement with the convergent
close-coupling calculations of Kheifets and Bray [Phys. Rev. A {\bf
58}, 4501 (1999)] but deviate notably from the experimental data of
Samson {\em et al.} [J. Phys. B {\bf 27} 887 (1994)].

\end{abstract}

\maketitle

\section{Introduction.}

Photoionization of two-electron atoms has been studied theoretically
by different authors starting from the pioneering paper by
\citet{wheel}. Review of  early literature on this subject
can be found in Ref.~\cite{amus}. Subsequently, a large number of
computations of helium photoionization cross-sections was reported
\cite{hh,sm93,ct,hin}. These calculations produced a collection of
results varying typically by  5\% from each other.  On the
experimental side, the benchmark set of data was reported by
\citet{sam} who measured the total photoionization cross-section of He
in the photon energy range from the threshold to 120 eV.  Agreement
between the theoretical and experimental data was within the same
margin of 5\%. In the following years, the theoretical interest
shifted towards calculation of differential characteristics of the
photoionization process and to studies of double photoionization.
Here, several approaches have been advocated including the many-body
perturbation theory \cite{hin,hin1}, convergent close-coupling method
\cite{KB98b,KB2000}, time-dependent close-coupling method
\cite{pr98,cp2002,cp2004}, $R$-matrix approach
\cite{mgb95,MSK00}, and methods based on the computation of
the dipole response function \cite{dalg} or $B$-spline implementations
of the exterior complex scaling \cite{horn}.

Due to this shift of focus, there have been no further attempt to
produce a consistent set of photoionization cross-sections of He with
an accuracy better than several percent. \citet{YSD98} combined
measurements of \citet{sam} at low energies and theoretical
calculations at high energies to construct a set of photoionization
cross sections of He that should be reliable at all energies. However,
there was no consistency check applied to the experimental data. In
the meantime, accurate helium photoionization cross-sections would be
highly desirable due to importance of He in astrophysics and its use
as a standard gas in determination of the photoionizaiton
cross-sections of other atomic and molecular species.

In the present paper, we develop the complex rotation method (CRM) for
highly accurate calculations of the total photoionization
cross-section of two-electron atomic targets.  One way of calculating
the photoionization cross-section is to combine the CRM technique with
the perturbation theory with respect to interaction of the atom with
the electromagnetic field. In such a perturbation theory, the CRM
 provides the basis of the field-free atomic states. It was
demonstrated by \citet{jr} that relying on the spectrum of the CRM
eigenvalues, one can construct a representation of the complete
Green's function of the atom. This, in turn, allows to write down a
convenient representation for the projection operator corresponding to
the continuous spectrum of the atom \cite{bgd}. Using this projection
operator, one can compute probabilities of transitions into continuum
under the action of some perturbation, in particular, the interaction
of the atom with the electromagnetic field.  Calculations of total
photoionization cross-sections of the helium atom based on this
technique have been reported in Refs.~\cite{gd,fh}. Similar ideas
were also used to determine static and dynamic polarizabilities of
helium \cite{pb92,bhd94}.

In the present work, we use the CRM procedure in a somewhat different,
non-perturbative way by applying it to the whole system the atom plus
the electromagnetic field. Thus, we are capable of going beyond the
perturbation theory and considering very strong fields.  In this
respect, the present technique has certain features in common with
Refs.~\cite{mn1,mn4}, where an approach based on the configuration
interaction procedure was developed for atoms with more than one
electron.  We cast our formalism using the language of
square-integrable functions with a finite norm.  This approach becomes
feasible in the so-called Floquet-Fourie representation of the
time-dependent Schr\"odinger equation.

Another key ingredient of the present work is the Hylleraas
basis functions which have long been used in various variational-type
calculations.  An excellent review of applications of the Hylleraas
basis to calculations of energies of two-electron atoms is given by
\citet{dr}. A well-known trademark of the Hylleraas basis set is a
very high accuracy of the atomic energies. In the present paper, we
show that the same high accuracy which is achieved for field-free
atomic states can also be attained when the atom is subjected to a
monochromatic electromagnetic field. In particular, the total
photoionization cross-sections can be calculated with an accuracy of
the order of a fraction of a percent.

Thus generated cross-sections were compared with the experimental
results of \citet{sam}.  We discovered a
systematic deviation from the experiment, especially in the region
close to double ionization threshold at the photon energies of
$\sim$80~eV. This deviation was confirmed by comparison with earlier
results produced by the convergent close-coupling (CCC) method
\cite{KB98d}.

The true potential of the present method is realized in the strong
field regime where the perturbation theory fails. As demonstrated
below, the Floquet-Fourie-Hylleraas ansatz produces very accurate
results in this regime as well.

\section{Theory}

\subsection{General Theory.}

The non-relativistic Hamiltonian of the helium atom in the presence of
the external monochromatic linearly-polarized electromagnetic field
can be written as :
\begin{equation}
\hat H=\hat T+\hat U +\hat V,
\label{ham}
\end{equation}
where, $\hat T$ is a kinetic term:
\begin{equation}
\hat{T}=
{{\bm p}_1^2\over 2}+
{{\bm p}_1^2\over 2},
\label{t}
\end{equation}
$\hat U$ potential energy term:
\begin{equation}
\hat{U}=-{2\over r_1}-{2\over r_2}+{1\over|{\bm r}_1-{\bm r}_2|},
\label{u}
\end{equation}
and $\hat V$ describes interaction of atom and the field. In the
length gauge (which will be used in the present paper), this
operator assumes the form:
\begin{equation}
\hat{V}={\bm F}\cdot{\bm D}\cos{\omega t},
\end{equation}
with $\displaystyle D={\bm r}_1+{\bm r}_2$.  Unless stated otherwise,
the atomic units are used throughout the paper.

We write the solution of the time-dependent Schrodinger equation
(TDSE) using the Floquet-Fourie ansatz \cite{BFJ91}
\begin{equation}
\Psi(t)=e^{-iEt}\sum\limits_n u_ne^{-inwt} \ .
\label{floq}
\end{equation}
By substituting this expression into the TDSE and equating
coefficients with $\displaystyle e^{-iEt-imwt}$, we obtain a chain
of coupled equations for the Floquet-Fourie coefficients $u_n$:
\begin{equation}
(E-\hat T -\hat U+n\omega)u_{n}=
{{\bm F}\cdot{\bm D}\over 2}(u_{n-1}+u_{n+1})
\ \ , \ \
n=0,\pm1\ldots,
\label{fl1}
\end{equation}
We solve this set of equations with the help of the complex rotation
procedure \cite{crot0,crot1,crot2,crot3,crotm,crotm1}.
Formally, the CMR can
be described as a complex transformation of radial variables
$\displaystyle r_i\to r_i e^{i\theta}$, where $\theta$ is the rotation
angle, the sole parameter defining the transformation.

Under this transformation, the chain of equations (\ref{fl1})
is converted into
\begin{equation}
(E-\hat T
e^{-2i\theta}-
\hat U e^{-i\theta}
+n\omega)u_{n}={{\bm F}\cdot{\bm D}\over 2}
e^{i\theta}(u_{n-1}+u_{n+1})
\ \ , \ \
n=0,\pm1\ldots,
\label{fl2}
\end{equation}
According to the general theory of CRM \cite{crot0,crot1,crot2,crot3},
the set of equations (\ref{fl2}) can be solved by means of variational
techniques if the rotation angle $\theta$ is properly chosen.

We introduce a basis set of square integrable functions $|n,k\rangle$
where the index $n$ refers to the number of the Floquet block and the
index $k$ denotes a particular $L^2$ function in the subspace of the
$n$-th block so that
$\displaystyle u_n=\sum\limits_{k} c_{nk}|n,k\rangle$.
With these notations, the set of Eqs.(\ref{fl2}) can be
rewritten in a matrix form as:
\begin{equation}
\sum\limits_k\langle n_1,k_1|E+n\omega-\hat T e^{-2i\theta}-
\hat U e^{-i\theta}|n,k\rangle c_{nk}=
\sum\limits_{n_2=n\pm 1,k}
\langle n_1,k_1|{{\bm F}\!\cdot\!{\bm D}\over 2}e^{i\theta}|
n_2k_2\rangle c_{n_2k}
\label{fl3},
\end{equation}
Notations can be further simplified by introducing obvious shorthands:
\begin{equation}
\left((E+n\omega)R_{n_1k_1}^{nk}-T_{n_1k_1}^{nk}e^{-2i\theta}-
U_{n_1k_1}^{nk}e^{-i\theta}\right)c_{nk}=
\sum\limits_{n2=n\pm 1} V_{n_1k_1}^{n_2k}{e^{i\theta}\over 2} c_{n_2k}
\label{fl5} ,
\end{equation}
where it is understood that summation is carried over the
repeated $k$-index.
Here $\displaystyle \hat V={\bm F}\!\cdot\!{\bm D}$, and $R$, $T$
and $U$ stand for the overlap, kinetic energy and potential energy
matrices, respectively.

One could say here a few words about the choice of the basis
allowed by the structure of the system (\ref{fl5}).
Suppose first, that in each of the subspaces corresponding to
different Floquet blocks we chose some compete set of functions,
such that for any $u_n$ in Eq.(\ref{fl2}) we had:
$\displaystyle u_n=\sum c_{nm} |n,m\rangle$.
Let the set of $|n,m\rangle$'s be
the same for all Floquet subspaces. Then, if we have retained $N$
Floquet blocks in the system (\ref{fl2}) and keep $M$ terms
in the expansion for each $u_n$ in Eq.(\ref{fl2}) we have
altogether $NM$ unknowns $c_{nm}$
in the system (\ref{fl2}). To get
a correctly posed eigenvalue problem, we should have the same
number of equations. This number is provided by projecting
each of the equations (\ref{fl2})
on one of the $|n,m\rangle$'s with $m=1\ldots M$.
This way of reducing the set of equations (\ref{fl2}) to
the form of matrix eigenvalue problem is correct, but
too general for our purposes. It can be seen, that
one can considerably diminish the resulting dimension of the
matrix eigenvalue problem by using certain symmetry properties
of the system Eq.(\ref{fl2}). It is easy to see, that
this system allows the following class of solutions:
$u_n$'s with even $n$ are of even parity, while $u_n$'s
with odd $n$ are of odd parity. Parity here is understood
with respect to the spatial inversion. Of course, there
is a class of solutions with the opposite property:
$u_n$'s with even $n$ are of odd parity, while $u_n$'s
with odd $n$ are of even parity. The solution we are looking
for (which is to describe behavior of the even $^1S^e$ state
of helium) evidently belongs to the first class.
We can therefore, choose the basis set as follows.

Instead of choosing the same set $|n,m\rangle$ for each
Floquet block, we choose two sets: a set
$|n_{\rm even},m\rangle$, consisting of basis finctions
of even parity, is used as a basis to represent $u_n$'s
with even $n$'s. Another set $|n_{\rm odd},m\rangle$,
composed of odd parity functions
is used as a basis to represent $u_n$'s with
odd $n$'s. Suppose that in the expansions of $u_n$'s with even $n$'s we
retain $M_{\rm even}$ terms, and in the expansions of $u_n$'s
with odd $n$'s - $M_{\rm odd}$ terms. Let the number of Floquet blocks
with even and odd $n$'s be respectively
$N_{\rm even}$ and $N_{\rm odd}$.
Than we have $N_{\rm even}M_{\rm even}+N_{\rm odd}M_{\rm odd}$
unknown coefficients
$c_{nm}$. We obtain the same number of equations by projecting
equations (\ref{fl2}) on $|n_{\rm even},m\rangle, m=1\ldots M_{\rm even}$
for even $n$
and on $|n_{\rm odd},m\rangle,m=1\ldots M_{\rm odd}$ for odd $n$.
Projection of
equations with even $n$ on the $|n_{\rm odd},m\rangle$ and
of equations with odd $n$ on the $|n_{\rm even},m\rangle$ gives identically zero
and does not add new equations. More details about the basis
functions $|n_{\rm even},m\rangle$ and $|n_{\rm odd},m\rangle$ is given below.

According to the general theory of CRM, some of the energy values
(generally complex) for which system~(\ref{fl5}) has a solution are
related to the position and width of the resonance state via $
E=E_r-i\Gamma/ 2$, where $E_r$ is position of the resonance and
$\Gamma$ its width. This leads one to solving a generalized eigenvalue
problem.  Effectiveness of  finding eigenvalues of such a
problem  depends crucially on the choice of the basis used
to represent the matrices in Eq.(\ref{fl5}).

\subsection{Basis set.}

The basis set used in the present paper was constructed
from the Hylleraas type functions:
\begin{equation}
g_{n_1,n_2,N}({\bm r}_1,{\bm r}_2)=
r^{n_1}_1 \ r^{n_2}_2 \ |{\bm r}_1-{\bm r}_2|^N e^{-ar_1-br_2}
|l_1(1)l_2(2)L\rangle,
\label{hil}
\end{equation}
where $a$,$b$ are some constants (to be specified below),
$n_1$,$n_2$,$N$ are integers and the angular part
\begin{equation}
|l_1(1)l_2(2)L\rangle=\sum\limits_{m_1m_2} C^{LM}_{l_1m_1l_2m_2}
Y_{l_1m_1}({\bm n}_1)
Y_{l_2m_2}({\bm n}_2),
\label{hil1}
\end{equation}
represents two angular momenta $l_1,l_2$ coupled to a state with a
total angular momentum $L$. The basis functions (\ref{hil}) must also
be properly symmetrized with respect to exchange of the electron
coordinates. When choosing parameters in \Eref{hil}, we followed the
following empirical rules \cite{dr,hil}. All the basis functions with
the parameters satisfying:
\begin{equation}
\label{max}
n_1+n_2+N< N_{\max}
\end{equation}
were included in the calculation. The parameter $N_{\rm max}$
determines the overall size of the basis. There is also a
semiempirical rule for choosing angular momenta $l_1$,$l_2$
in the Eq.(\ref{hil}). Thus, for states of the natural
parity $l_1$,$l_2$ are best chosen so that $l_1+l_2=L$.
Both these criteria help to avoid the numerical
problems due to near-degeneracy of the basis set when
its dimension becomes large.

\section{Numerical Results}

\subsection{Field-free case}

In the present work, our main goal is to obtain accurate
photoionization cross-sections from the ground state of neutral helium
for not very large electromagnetic field intensities.  Accordingly,
our main interest is focused on the states of $S$ and $P$ symmetries.
Therefore, our first goal is to choose such a basis that solution of
the eigenvalue problem (\ref{fl5}) yields accurate energies for the
ground $^1S$ and first excited $^1P^o$ state of the helium atom in the
absence of the field.

This goal was achieved as follows.  We chose parameters $N_{\rm
max}=18$, $a=b=2$ for the $S$-states and $N_{\rm max}=13$, $a,b=1,2$
for the $P$-states. The reason for enlarging the basis set for the
excited $P$-states is that the electrons in such states are generally
on different distances from the nucleus. This choice combined with
restriction on angular momenta \eref{max} resulted in $N_S=372$ basis
functions for the $S$-states and $N_P=660$ basis functions for the
$P$-states.

The next step was to solve the generalized eigenvalue problem for the
field-free case.  In \Eref{fl5} we put $F=0$, $\omega=0$, and
limited ourselves to the blocks with $n=0$, $n=\pm 1$, the $n=0$ block
being composed of the states of $^1S^e$ symmetry, and $n=\pm 1$ blocks
composed of the states of $^1P^o$ symmetry.  All the numerical results
reported below were obtained using the quadruple precision
arithmetics.

We note, that in the presence of the weak electromagnetic field
account of the blocks with $n=\pm 1$ corresponds to absorption and
emission of one photon. We shall use this fact below to extract the
photoionization cross-section from our calculation. For the moment, we
are concerned with testing the accuracy of our basis. Diagonalization
of the eigenvalue problem (\ref{fl5}) with $F=0$, $\omega=0$ in the
basis described above produced the following results for the complex
energies: $E=-2.903724384 + i \ 1.3\times 10^{-8}$ (the ground state)
and $E=-2.123843094+i \ 7.6\times 10^{-9}$ $ (1s2p ^1P^o$ state). A
small imaginary part which, in the absence of the field, should be
zero could be taken as an indication of an accuracy of our basis set.
Either this criteria or a direct comparison with the well-known
results of highly accurate calculations \cite{dr} shows that we have
achieved an accuracy of the order of $10^{-8}$ a.u. This accuracy, as
will be demonstrated below, is sufficient to obtain the
photoionization cross-sections with at least three significant figures.

\subsection{Total photoionization cross sections}

To calculate the total photoionization cross sections we adopted the
following strategy.  The eigenvalue problem (\ref{fl5}) was solved
with the Floquet blocks $n=0,\pm 1$ retained, the composition of each
block was the same as described above for the field-free case.
Diagonalization of the eigenvalue problem (\ref{fl5}) produced energy
shift and total width for the ground state. By definition, the
photoionization cross-section from this state is related to the total
width $\Gamma$ via
\begin{equation}
\sigma=\lim_{F\to 0} 8\pi \alpha \Gamma \omega/ F^2,
\label{sig}
\end{equation}
where $F$ is field strength, $\omega$ its frequency, $\alpha$ is the
fine structure constant.  We need therefore to extract from our
calculation the coefficient with $F^2$ in the asymptotic law defining
the weak-field behavior of the width:
\begin{equation}
\Gamma(F)=\Gamma_0 F^2+\Gamma_1F^3+\ldots
\label{as}
\end{equation}
To implement this strategy, we need an extrapolation procedure since
the calculation based on the system (\ref{fl5}) is performed for a
non-zero field strength. Although finite, this field strength should
not be too small to compute $\Gamma$ with sufficient accuracy.

The issue of accuracy can be addressed as usual in variational-type
calculations, by merely increasing the basis size and verifying that
the results do not change appreciably. Such a test was performed for a
photon energy $\omega=80$ eV and a field strength $F=0.1$ a.u.  by
varying the parameter $N_{\rm max}$ in Eq.(\ref{max}) for the $S$ and
$P$ states. The diagonalization of the problem (\ref{fl5}) was
performed with the Floquet blocks $n=0,\pm 1$ retained. All the
remaining details of the basis (nonlinear parameters etc.)  were the
same as in the field-free case reported above. The calculation was
performed for the value of the rotation angle $\theta=0.3$.

\begin{table}
\caption{\label{tabm1}
Results for the ground state eigenvalue of problem
(\ref{fl5}) as functions of parameters $N_{\rm max}$ in Eq.(\ref{max}),
$\omega=80$ eV, $F=0.1$ a.u.}
\begin{ruledtabular}
\begin{tabular}{ccccc}
\noalign{\smallskip}
$N_{\rm max}^S$ & $N_{\rm max}^P$ & Total dimension of the &
${\rm Re} E$(a.u.) &$\Gamma$ (a.u.) \\
&& eigenvalue problem (\ref{fl5}) & & \\
\noalign{\smallskip}
\hline
\noalign{\smallskip}
17 & 11 & 1300 & -2.90307660 & 0.000487738 \\
18 & 12 & 1692 & -2.90307661 & 0.000487698 \\
19 & 13 & 2204 & -2.90307659 & 0.000487689 \\
\end{tabular}
\end{ruledtabular}
\end{table}

The test results are presented in Table \ref{tabm1}. One can observe
that, just as in the field-free case, the accuracy is on the level of
$10^{-8}$ a.u., which implies that $\Gamma$ has at least four
significant digits in this interval of field strengths.

The issue of the stability of the results with respect to the number
of the Floquet blocks included in diagonalization of (\ref{fl5}) is
addressed in the next section where we consider effects of going
beyond the first order perturbation theory. We shall say in advance
that including the Floquet blocks with $n=\pm 2$ in diagonalization of
(\ref{fl5}) does not alter the numerical accuracy appreciably.

As to the extrapolation procedure needed to extract the coefficient
$\Gamma_0$ in Eq.(\ref{as}), we chose a scheme based on the
three-point Lagrange formula.  For each frequency reported below, we
performed calculations for the field strengths $F=0.07,0.1,0.13$a.u.
We also used a mid size basis set with $N_{\rm max}^S=18$, $N_{\rm
max}^P=12$, Floquet blocks with $n=0,\pm 1$, all other details of the
basis are the same as in the field-free case above.   Results of
this calculation and extrapolation are shown in Table \ref{tab2}.

\begin{table}
\caption{\label{tab2}
Extrapolation of the $\Gamma$'s to the zero-field limit.}
\begin{ruledtabular}
\begin{tabular}{lcccc}
&\multicolumn{4}{c}{{$\Gamma/ F^2$} (a.u.)} \\
\noalign{\smallskip}
$\omega$ (eV)  & $F=0.07$ a.u. & $F=0.1$ a.u. & $F=0.13$ a.u. & $F=0$ (Extrapolation) \\
\hline
\noalign{\smallskip}
40 & 0.4208622& 0.4201601& 0.4192063& 0.4215215  \\
80 & 0.0488002& 0.0487698& 0.0487239& 0.0488112  \\
85 & 0.0392854& 0.0392618& 0.0392330& 0.0393202  \\
91 & 0.0306858& 0.0306720& 0.0306524& 0.0306961  \\
95 & 0.0262180& 0.0262082& 0.0261936& 0.0262224  \\
111& 0.0147116& 0.0147084& 0.0147033& 0.0147116  \\
205& 0.0013719& 0.0013726& 0.0013729& 0.0013687  \\
\end{tabular}
\end{ruledtabular}
\end{table}

Using an estimate for the remainder of the series (\ref{as}), it is a
simple matter to verify that for the field strengths considered the
possible relative error introduced by the extrapolation of $ \Gamma/
F^2$ is of the order of 0.1\%. Hence, at least three digits in our
result for the extrapolated ratio $ \Gamma/ F^2$ and the
cross-sections reported below must be reliable.  This level of
accuracy can easily be improved by merely going to extrapolation
schemes of higher order and computing $\Gamma$ for more field values.

\begin{table}
\caption{\label{tab3}
Comparison of the present results and other theoretical and
experimental data for the total photoionization cross section (in
Mb).}
\begin{ruledtabular}
\begin{tabular}{lc cccc c }
$\omega$ (eV)  & Present & \multicolumn{4}{c}{CCC}  & Experiment  \\
               &         & L&V&A&Average&\citet{sam}\\
\hline
\noalign{\smallskip}
40 &  3.1822 &3.188   &  3.178   & 3.247  & 3.2043& 3.16  \\
80 &  0.7369 &0.7432  &  0.7403  & 0.7366 & 0.7400& 0.693 \\
85 &  0.6308 &0.6364  &  0.6327  & 0.6294 & 0.6328& 0.595 \\
91 &  0.5272 &0.5333  &  0.5284  & 0.5248 & 0.5288& 0.502 \\
95 &  0.4701 &0.4765  &  0.4717  & 0.4689 & 0.4723& 0.450 \\
111&  0.3082 &0.3097  &  0.3089  & 0.3081 & 0.3089& 0.300 \\
205&  0.0529 &0.0533 & 0.0534 & 0.0531  &0.0533 & 0.0510 \\
\end{tabular}
\end{ruledtabular}
\end{table}
\bigskip

In Table \ref{tab3} we present our results for the cross-sections
based on formula (\ref{sig}) in which we fed the extrapolated ratios
from the last column of Table \ref{tab2}. Along with our data, we
present the benchmark experimental results of \citet{sam} as well as
earlier theoretical results from Ref.~\cite{KB98d}.  The experimental
setup of \citet{sam} was such that the measured cross-section was
summed over all final states of the remaining ion including the
doubly ionized states. It is exactly the cross-section that is
calculated presently and therefore  comparison between the theory
and experiment should be straightforward.

\begin{figure}[h]
\epsfxsize = 16cm
\epsffile{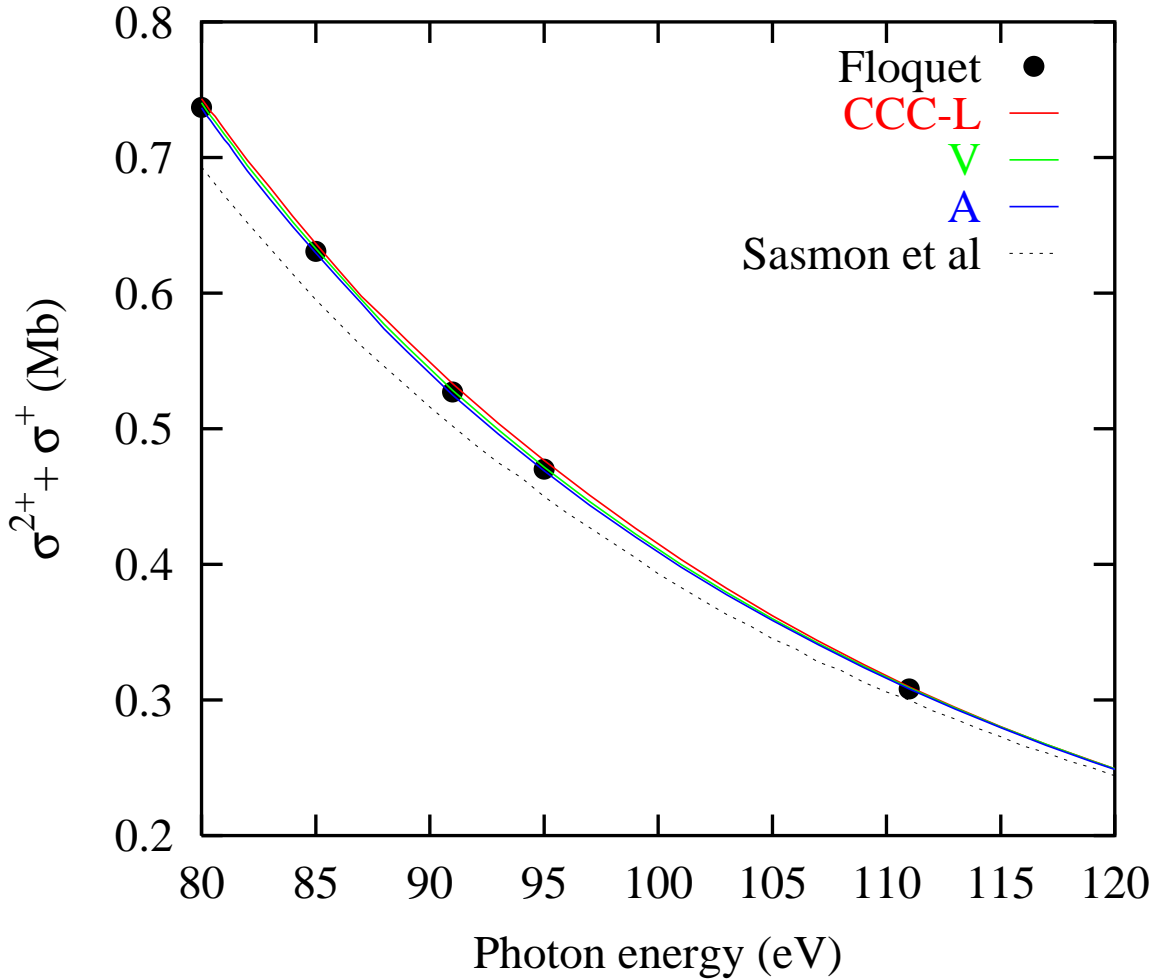}
\caption{\label{fig1}
The total photoionization cross-section (sum of the single $\sigma^+$
and double $\sigma^{2+}$ cross-sections) as a function of the photon
energy. The present calculation for selected photon energies is
denoted by dotes. The CCC calculation in the three gauges of the
electromagnetic interaction (Length, Velocity and Acceleration ) is
exhibited by  different colors / line styles. The experiment of
\citet{sam} is presented by a dotted line.
}
\end{figure}

Our theoretical results agree with the data of \citet{sam} within the
postulated experimental accuracy of few percents.  The strongest
deviation is for $\omega=80$~eV where the difference between the
present result and the experimental value is 6\%. This is deviation is
clearly seen in the Figure where we plot the present Floquet
calculation along with the CCC calculation in three gauges of the
lectromagnetic interaction and the experiment.  Agreement between the
present calculation and that of the CCC is much better,
difference of the results of two approaches not exceeding
1\%. The accuracy of the CCC result is hard to estimate directly as
this method relies on the numerical solution of a set of
close-coupling equations. The only implicit indication is the
difference between the cross-sections calculated in the three gauges
of the electromagnetic interaction, the length (L), velocity (V) and
acceleration (A). This difference is typically 1-2\%. Thus, the
deviation of the present calculation with the CCC is more likely to be
the problem of the latter as the former is believed to be much more
accurate.

As a by-product of the calculation described above, we
also obtained the shift of the ground state of helium due to
the interaction of atom with the linearly polarized monochromatic
field (Table \ref{tabs}).

\begin{table}
\caption{\label{tabs}
Real part of the energy of the ground state of helium in the presence
of the linearly polarized monochromatic field.}
\begin{ruledtabular}
\begin{tabular}{lccc}
\noalign{\smallskip}
&\multicolumn{3}{c}{{${\rm Re} E$} (a.u.)} \\
\noalign{\smallskip}
$\omega$ (eV)  & $F=0.07$ a.u. & $F=0.1$ a.u. & $F=0.13$ a.u. \\
\hline
\noalign{\smallskip}
40 & -2.90281954& -2.90187690& -2.90060016 \\
80 & -2.90340686& -2.90307658& -2.90263014 \\
85 & -2.90344158& -2.90314741& -2.90274973 \\
91 & -2.90347684& -2.90321933& -2.90287117 \\
95 & -2.90349708& -2.90326062& -2.90294089 \\
111& -2.90355845& -2.90338582& -2.90315235 \\
205& -2.90367802& -2.90362977& -2.90356450  \\
\end{tabular}
\end{ruledtabular}
\end{table}

\newpage
\subsection{Extended calculation}

We now turn to extended calculations with inclusion of a larger
number of the Floquet blocks $n=0,\pm 1, \pm 2$ in \Eref{fl5}.  The
aim of these calculations is two-fold. First, we shall confirm the
stated accuracy of the present weak field results which is not
effected by the number of the Floquet blocks retained in the
calculation.  Second, we report some preliminary results concerning
behavior of the widths parameter in stronger fields where inclusion
of a larger number of the Floquet blocks becomes essential due to a
non-perturbative nature of the processes involved.

The basis for the extended calculations was constructed as
follows.
As we discussed above
the basis subset spanning each Floquet block in the system
(\ref{fl5}) can be chosen
to consist of the functions of a given parity, two adjacent blocks
having opposite parities. Thus, in
the low-field calculations described above, the block $n=0$ was
composed of even basis functions while two blocks with $n=\pm 1$
contained odd basis functions.  Inclusion of the blocks with $n=\pm 2$
is, therefore, equivalent to adding more even basis functions. We did
it in the following way.  In addition to the $^1S^e$ states we
previously had in the $n=0$ block, the states of the symmetries
$^1D^e$ and $^1P^e$ were included in the calculation. Thus the blocks
with $n=0$ and $n=\pm 2$ had the following composition: $N_{\rm
max}=18$ for the $^1S^e$-basis functions, $N_{\rm max}=8$ for the
$^1P^e$ and $^1D^e$-basis functions. As before, the blocks with $n=\pm
1$ were composed of basis functions of $^1P^o$-symmetry with $N_{\rm
max}=13$. Thus, the basis set is considerably enlarged comparing to
the one used in the previous section. With this choice of parameters
$N_{\rm max}$, the overall dimension of the eigenvalue problem
(\ref{fl5}) was 2676. Results produced for the ground state of He by
diagonising this eigenvalue problem are shown in Table \ref{taben}

\begin{table}
\caption{\label{taben}
Results of the calculation with Floquet blocks $n=0,\pm 1,\pm 2$
included in the system (\ref{fl5}).}
\begin{ruledtabular}
\begin{tabular}{lcccc}
\noalign{\smallskip}
&\multicolumn{2}{c}{$\omega=111$ eV}&
\multicolumn{2}{c}{$\omega=205$ eV} \\
\noalign{\smallskip}
$F$ (a.u.)  & ${\rm Re} E$ (a.u.) & $\Gamma/ F^2$ (a.u.)
& ${\rm Re} E$ (a.u.) & $\Gamma/ F^2$ (a.u.) \\
\hline
\noalign{\smallskip}
0.10 & -2.90338569 & 0.014714 & -2.90362976& 0.0013734  \\
0.13 & -2.90315198 & 0.014715 & -2.90356447& 0.0013739  \\
0.20 & -2.90236955 & 0.014715 & -2.90334589& 0.0013743  \\
0.50 & -2.89525524 & 0.014706 & -2.90135842& 0.0013745  \\
1.0  & -2.86985102 & 0.014665 & -2.89426254& 0.0013701  \\
\end{tabular}
\end{ruledtabular}
\end{table}

Comparison of the results given in Table \ref{taben} supports the
assertion we made in the previous section as to the accuracy of our
results for the widths.
As one can see, for the field strengths $F\approx 0.1$ a.u.,
inclusion of the additional Floquet blocks
and basis states of symmetries other than $S$ and $P$ produces
relative variations in the widths on the order of $0.01$
percent. This means that for such field values 
we are still within the domain of the validity of
the perturbation
expansion. For the frequencies presented in the Table the domain
of the perturbation theory actually extends quite far in the
region of large field strengths. As one can see from the Table
(\ref{taben}), the ratio $\Gamma/F^2$ starts changing in a
more or less appreciable manner only for field strengths as large
as $F\approx 1$ a.u.

\section{Conclusion}

We performed a calculation of the total photoionization cross-sections
from the ground state of helium.  We employed a theoretical procedure
based on the Floquet-Fourie representation of the solutions of the
TDSE describing the helium atom in the presence of the linearly
polarized monochromatic electromagnetic field. The resulting set of
Floquet equations was cast into a generalized eigenvalue problem by
the complex rotation method. Our approach is essentially
non-perturbative.  This is in contrast with other works \cite{bgd,fh}
where the CRM was used to produce an accurate description of the
field-free helium atom, thus giving the ground for application of the
perturbation theory. In our approach, we do not rely on any
perturbation expansion to describe interaction of the atom and the
electromagnetic field. This interaction is included into the theory
from the beginning.  We would like to emphasize the accuracy of the
present results for the photoionization cross-sections which, we
believe, is on the level of a fraction of a percent. Although only few
selected photon energies were reported in the paper, far wider and
denser energy grid was covered by the present calculation. These
results might serve as an accurate database and find their use in
various astrophysics and atomic physics applications. The authors
shall gladly communicate these data on request.

\section{Acknowledgements}
The authors acknowledge support of the Australian Research Council in
the form of  Discovery grant DP0451211.

\end{document}